\documentclass[aps,10pt,letterpaper,twocolumn,english,showpacs,preprintnumbers,amsmath,amssymb,nofootinbib,prl]{revtex4-1}
\usepackage{amsmath}
\usepackage{graphicx}
\usepackage{setspace}
\usepackage{amssymb}
\usepackage{dcolumn}
\usepackage{mathrsfs}
\usepackage{bm}

\begin{document}

\preprint{}
\title{Tomography of the temporal-spectral state of subnatural-linewidth single photons from atomic ensembles}

\author{Ce Yang,$^{1}$ Zhenjie Gu,$^{1}$, Peng Chen,$^{1}$ Zhongzhong Qin,$^{3,4}$ J. F. Chen,$^{1,3,\dagger}$ and Weiping Zhang$^{2,3,\dagger}$}
\affiliation{$^1$Quantum Institute for Light and Atoms, Department of Physics, East China Normal University, Shanghai 200241, China\\
$^2$School of Physics and Astronomy, and Tsung-Dao Lee Institute, Shanghai Jiao Tong University, Shanghai 200240, China\\
$^3$Collaborative Innovation Center of Extreme Optics, Shanxi University,
Taiyuan, Shanxi 030006, China\\
$^4$State Key Laboratory of Quantum Optics and Quantum Optics Devices, Institute of Opto-Electronics, Shanxi University, Taiyuan 030006, People¡¯s Republic of China\\
$^{\dagger}$e-mail: jfchen@phy.ecnu.edu.cn; wpzhang@phy.ecnu.edu.cn}

\date{\today}

\begin{abstract}
Subnatural-linewidth single-photon source is a potential candidate for exploring the time degree of freedom in photonic quantum information science. This type of single-photon source has been demonstrated to be generated and reshaped in atomic ensembles without any external cavity or filter, and is typically characterized through photon-counting technology. However, the full complex temporal mode function(TMF) of the photon source is not able to be revealed from direct photon counting measurement. Here, for the first time, we demonstrate the complete temporal mode of the subnatural-linewidth single photons generated from a cold atomic cloud. Through heterodyne detection between the single photon and a local oscillator with various central frequencies, we recover the temporal density matrix of the single photons at resolvable time bins. Further we demonstrate that the reduced autocorrelation function measured through homodyne detection perfectly reveals the pure temporal-spectral state of the subnatural-linewidth single photons.
\end{abstract}


\maketitle

\begin{figure*}
\includegraphics[width=12cm]{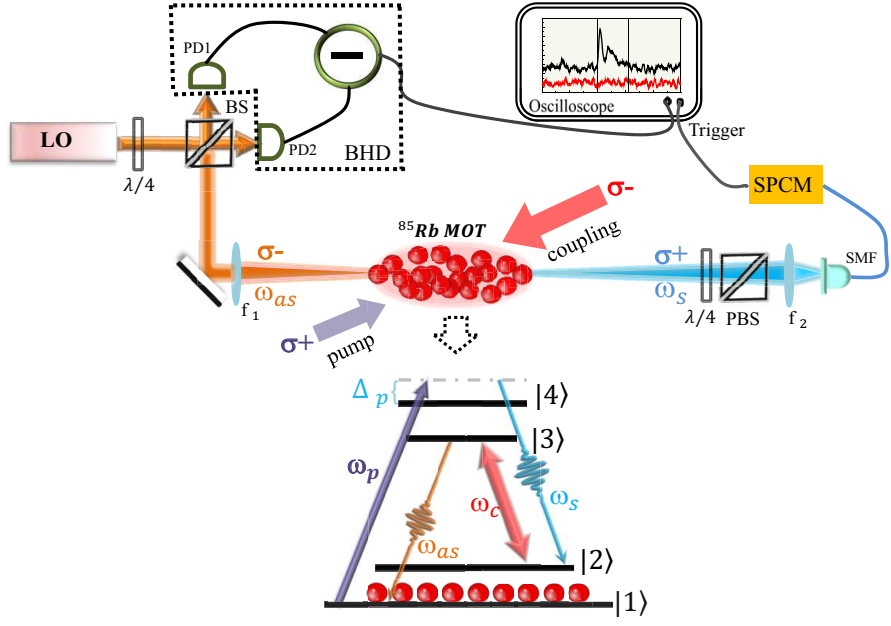}
\centering
\caption{Schematics of optical setup and atomic energy level. The paired Stokes and Anti-Stokes photons are generated from SFWM process induced by two classical laser beams (pump and coupling) in a $^{85}$Rb cold atom cloud. The Stokes photons are coupled into single-mode fibers (SMF) for photon counting. The Anti-Stokes photons are directed to a 50/50 beam splitter (BS). A continuous laser with power of 18 mW, phase locked with the coupling laser, is served as LO. BHD: balanced homodyne detector. SPCM: single photon counting module. In the double-$\Lambda$ energy level scheme, $|1\rangle=5S_{1/2},F=2$, $|2\rangle=5S_{1/2},F=3$, $|3\rangle=5P_{1/2},F=3$ and $|4\rangle=5P_{3/2},F=3$. The coupling laser with Rabi frequency $\Omega_c$ is on resonant to the transition $|2\rangle\rightarrow|3\rangle$, and constitute an EIT energy level configuration for the Anti-Stokes photons.}
\label{fig:experiment}
\end{figure*}

In thriving photonic quantum technologies, single photons are one of the most important carriers with quantum states. Qubits or qudits (states with more than two dimensions) can be encoded in various degrees of freedom, e.g., polarizations, spatial and temporal-spectral modes. Compared to other degrees of freedom, encoding information in temporal-spectral modes are favorable in long-distance communication through fibers\cite{Marcikic,Tapster}. Actually time-bin qubits\cite{Franson,Kwiat} are widely used in quantum teleportation\cite{Thew,Riedmatten,Marcikic,Jin}, quantum key distribution\cite{IslamScience,Martin} and other quantum communication protocols\cite{Brendel,Bimbard,Gorshkov}. Conventional scheme to generate time-bin qubits is to split a single photon into two separate time bins through an unbalanced interferometer, i.e, Franson interferometer. The system is compact but hard to be extended to multi-dimensions, since every splitting will decrease the photon rate dramatically.

In recent years, utilizing complete and orthogonal temporal-spectral modes based on the ultrafast photons produced from spontaneous parametric down conversion (SPDC) is introduced\cite{RaymerPRX} and developing fast both theoretically and experimentally\cite{CKLaw,Brecht,Ansari}. However, due to the ultra-broad-band feature of these PDC photons, increasing the spectral modes will inevitably reduce the temporal purity of the single-photon state\cite{Ansari}, which is essential in quantum information processing. Following this idea, it is possible to encode quantum information directly in the temporal mode of a single photon with ultra-long coherence time\cite{Zhiguang,LuweiPRA}. The purity of the temporal reshaped single photon is preserved to be near unity as long as detectors' time resolution is below 10\% of the coherence time\cite{Dutheory,Qian}.

The sub-linewidth biphotons and single photons are demonstrated to be produced from atomic ensembles\cite{Balic,DuPRL,CShu}. With electromagnetically-induced transparency (EIT)\cite{EIT}, the temporal length of the biphotons can be tailored from nanoseconds to microseconds, which shows great advantage in single photon waveform manipulation\cite{ChinmayPRA,Kolchin,Shanchao}. The EIT effect provides an additional route for the temporal mode of these narrow-band nonclassical light source to be manipulated with arbitrary waveforms\cite{JFChen,Luweispatial,YoungWook}. In these reports of photon-source, filters and cavities are abandoned to avoid a pre-defined shape in spectral mode. So far, all of the temporal-spectral mode measurements on these photons are performed by photon counting technique which have no requirement of the photon mode. Therefore the amplitude of the mode function can be given directly, and the phase information can only be measured through certain indirect photon interference\cite{PChenPRL}. To simultaneously give the complete mode function of the photon source, homodyne detection which utilizes photon-current detection is a state-of art technique\cite{Smithey,Leonhardt,RevModPhys,MacRae,Bimbard,Qin}, which demands stringent mode-matching between the target photons and a laser beam served as the local oscillator (LO).


In this letter, for the first time, through matching the spatial mode of LO with the photon source, we demonstrate the complete temporal mode of the subnatural-linewidth single photons. The single photons are generated from spontaneous four-wave mixing process in a dense cold atom cloud. Through heterodyne detections between the single photon and LO with multiple central frequencies, we recover the temporal density matrix of the single photons at resolvable time bins. We therefore directly obtain a temporal purity over 90\% for the subnatural-linewidth EIT shaped single photons. Furthermore, we demonstrate that homodyne detection, whose LO utilizes the identical central frequency with the photon source, is sufficient to reveal the entire photonic temporal-spectral state.


We define the temporal mode function(TMF) in the frame of heralded single photons, which is obtained by heralding a partner photon from a photon pair. Assume that we have a pair of frequency-anti-correlated photon pair, and the two-photon state is denoted as,

\begin{equation}
|1_1,1_2\rangle_{\Phi}=\int d\Omega \Phi(\Omega)\hat{a}_2^{\dag}(\omega_{20}-\Omega)\hat{a}_1^{\dag}(\omega_{10}+\Omega)|0\rangle
\label{eq:biphoton state}
\end{equation}

\noindent where $|0\rangle$ denotes the vacuum, $\omega_{20}$ and $\omega_{10}$ correspond to the central angular frequencies of photons $1$ and $2$. $\hat{a}_1$ and $\hat{a}_2$ are their corresponding annihilation operators. $\Phi(\Omega)$ is the two-photon joint spectrum function. After photon 1 annihilated at time $t=0$ with its uncertainty infinitesimally small, the heralded single-photon state is expressed as a superposition state,

\begin{eqnarray}
|1\rangle_{\Phi}&=&\hat{a}_1(t)|1_1,1_2\rangle_{\Phi}  \\
&=&\frac{1}{\sqrt{2\pi}}\int d\Omega \Phi(\Omega)\hat{a}_2^{\dag}(\omega_{20}-\Omega)e^{i(\omega_{10}+\Omega)t}|0\rangle\nonumber\\
&=&\frac{1}{\sqrt{2\pi}}\int d\Omega \Phi(\Omega)\hat{a}_2^{\dag}(\omega_{20}-\Omega)|0\rangle\
\label{eq:heralded single state frequency}
\end{eqnarray}

\noindent Equivalently, the heralded single-photon state can be expressed as a superposition state of relative arrival times of $\tau$,

\begin{align}
|1\rangle_{\Phi}=\int d\tau \varphi(\tau)\hat{a}_2^{\dag}(\tau)e^{-i\omega_{20}\tau}|0\rangle
\label{eq:heralded single state time}
\end{align}

\noindent where the TMF $\varphi(\tau)=\frac{1}{\sqrt{2\pi}}\int d\Omega \Phi(\Omega)e^{-i\Omega\tau}$ is the Fourier transform of the joint spectrum function, and normally this TMF is a complex function with magnitude $|\varphi(\tau)|$ and phase $\theta(\tau)=arctan[\frac{Im(\varphi(\tau))}{Re(\varphi(\tau))}]$.

The above state can be expressed in a discrete time bin basis. We denote every resolvable $i$th time bin as $\tau_i=i\delta\tau$, where $\delta\tau$ denotes the detector's resolution. From $[\hat{a}(\tau_i),\hat{a}^{\dag}(\tau_j)]=\delta_{ij}$, $|1_i\rangle=|0_{\tau_1},0_{\tau_2},...,1_{\tau_i},0_{\tau_{i+1}},...\rangle$, and it denotes a single photon found in a specific time bin. In this time bin basis, the temporal density operator is defined as $\hat{\rho}_{TM}=\sum\limits_{ij} \rho_{ij}|1_j\rangle\langle1_i|$. For subnatural-linewidth single photons, the temporal density operator can be constructed from a pure state,

\begin{eqnarray}
\hat{\rho}_{TM}&=&|1\rangle_{\Phi \Phi}\langle1| \nonumber\\
&=&\sum\limits_{i}\sum\limits_{j}\varphi^{\ast}(\tau_i)\varphi(\tau_j)|1_{\tau_j}\rangle\langle1_{\tau_i}|
\label{eq:density operator in pure state}
\end{eqnarray}

\noindent In discrete time-bin basis, the matrix elements $\rho_{ij}=\varphi^{\ast}(\tau_i)\varphi(\tau_j)$. Therefore the characterization of the TMF of single photons can be realized by reconstructing the temporal density matrix.

\begin{figure*}
\centering
\includegraphics[width=18cm]{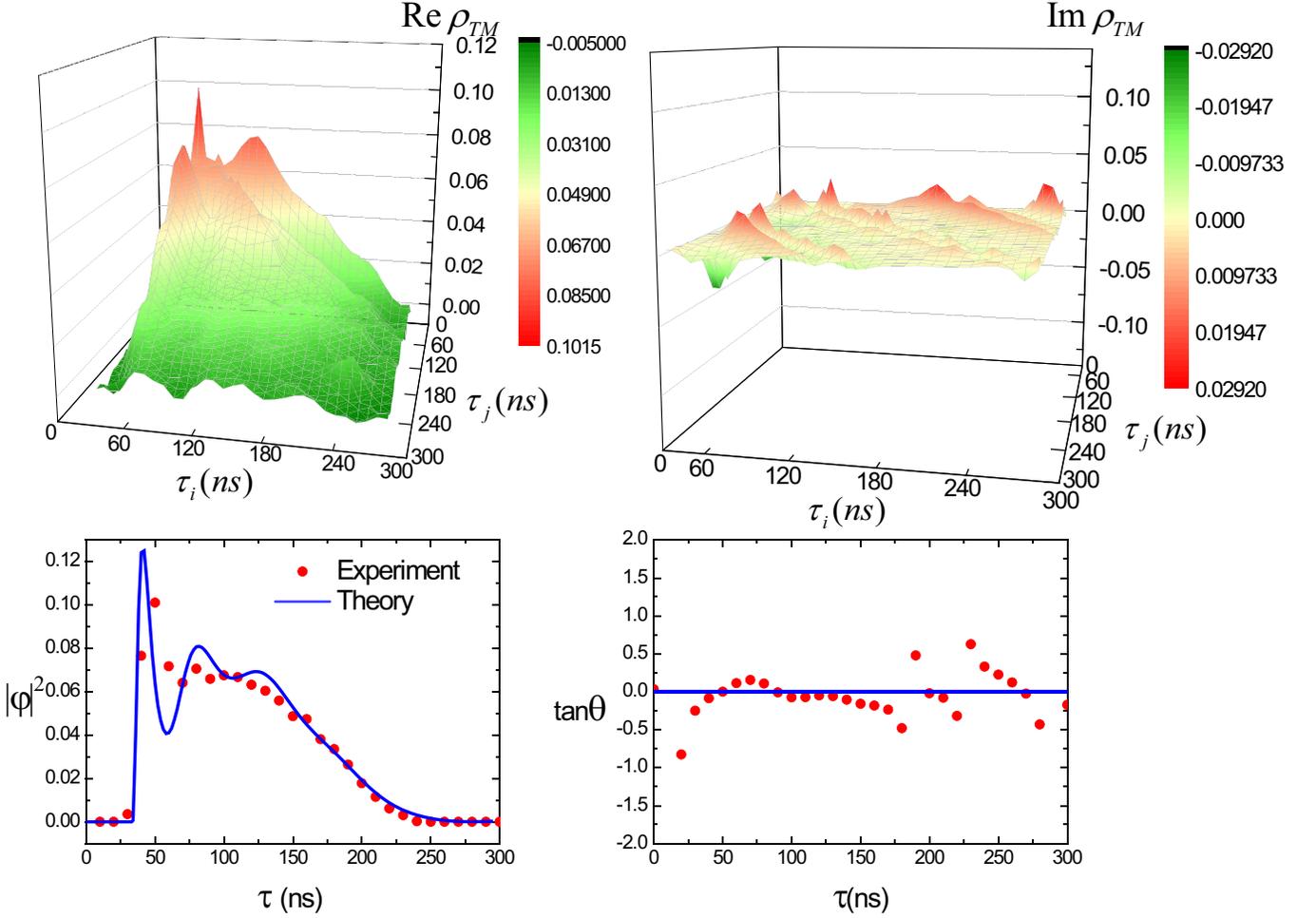}
\caption{Temporal mode characterization of subnatural-linewidth single photons. (a) Real part of the normalized temporal density matrix, $Re\rho_{TM}$; (b) Imaginary part of the normalized temporal density matrix, $Im\rho_{TM}$; (c) $|\varphi|^2$, obtained from the diagonal elements at $\tau_i=\tau_j$; (d) $Tan\theta$, obtained from $Im\rho_{mj}/Re\rho_{mj}$ where $m=50$. Here the value of $m$ is chosen so that $\rho_{mj}$ cross the maximum value of the matrix. In the SFWM process where subnatural-linewidth single photons are generated, $\Omega_c=2\pi\times18$ MHz, $OD=53$ . Totally $5\times10^5$ samples of photo-currents are taken to obtain the results. The time resolution of BHD used in this measurement is 30 ns.}
\label{fig:rho matrix}
\end{figure*}

The schematics of measuring the temporal mode of single photons is shown in Fig.~\ref{fig:experiment}. A pair of spectral anti-correlated photons, Stokes and Anti-Stokes, are generated from spontaneous four-wave mixing (SFWM) process in a two-dimensional $^{85}Rb$ magneto-optical trap (MOT)\cite{ShanchaoRSI,Jun}. Two continuous-wave (cw) lasers in the SFWM are, a 780nm pump which is blue-detuned from $|1\rangle$ to $|4\rangle $ by $146$ MHz, and a 795nm coupling resonant with $|2\rangle\rightarrow |3\rangle$. They are applied onto the atom cloud simultaneously in a backward scattering configuration, and the pump-coupling axis is deviated from the longitudinal axis of the 2D MOT by $\theta=2.5^\circ$. The pump and coupling beams are circularly polarized, corresponding to $\sigma^+$ and $\sigma^-$ respectively, and therefore the induced Stokes and Anti-Stokes photons are collected with $\sigma^+$ and $\sigma^-$, respectively. To ensure a low excitation rate of SFWM, the Rabi frequency of the pump beam is set as $\Omega_p=2\pi\times3.3$ MHz and the count rates of Stokes and Anti-Stokes photons are 4400/s and 4000/s, respectively. We utilize a continuous 795 nm laser beam which is phase locked with the coupling laser of the SFWM process, as LO to interfere with the Anti-Stokes photons. The polarization and spatial mode of LO are carefully matched with those of the Anti-Stokes, which is mimic with a visible laser beam emitted from the single-mode fiber (SMF) collecting the Stokes. The central frequency of LO is tunable around $\omega_{as}$ with an acoustic-optical modulator (AOM, Brimose). The photo-current of the BHD is proportional to the quadrature $\hat{X}=(\hat{a}e^{-i\theta}+\hat{a}^{\dag}e^{i\theta})/\sqrt{2}$ of the Anti-Stokes. For heralded single-photon source, the quadrature should be measured with triggers of the partner photons. In our setup, we collect the Stokes photons with a SMF and the trigger is given whenever the single-photon counting module (SPCM) detects a photon.  Therefore, we obtain $\langle\hat{X}(\tau)\rangle$ from traces of currents through a high-speed digital oscilloscope trigged by the Stokes photons.

The correlation between different time bins can be expressed as $\langle X_iX_j\rangle=\delta_{ij}/2+A_{ij}$, where $\delta_{ij}/2$ ($\delta_{ij}=1$ only when $i=j$) denotes the autocorrelation matrix for the vacuum. $A_{ij}$ is the reduced autocorrelation matrix for the Anti-Stokes photons, with trigger of a Stokes photon, and is related to the temporal density matrix as\cite{Qin},

\begin{eqnarray}
A_{ij}=&Re[\rho_{TM,ij}]cos[\Delta\omega(t_i-t_j)]\nonumber \\
&+Im[\rho_{TM,ij}]sin[\Delta\omega(t_i-t_j)]
\label{eq:Aij}
\end{eqnarray}

\noindent $\Delta\omega$ is the frequency difference between LO and the Anti-Stokes photons. With balanced homodyne detection, i.e., $\Delta\omega=0$, $Re[\rho_{TM,ij}]$ is obtained directly from the autocorrelation matrix. With heterodyne detection, i.e., $\Delta\omega\neq0$, $Im[\rho_{TM,ij}]$ can be obtained from two pairs of $A_{ij}$ and $\Delta\omega$. To minimize statistical uncertainty, we choose 8 sets of LO frequencies, which are equivalent to $\Delta\omega$$=-10$, $-5$, $0$, $3$, $8$, $13$, $18$, $23$ MHz, to calculate the real and imaginary part of the temporal density matrix. Fig.\ref{fig:rho matrix}(a) shows the normalized $Re[\rho_{TM}]$, which is calculated through the autocorrelation matrix $A_{ij}$ when $\Delta\omega=0$. The diagonal elements $i=j$ are obtained through minimizing a cost function, while obeying the normalization condition $Tr(\rho)=1$. Here, the cost function is taken as the difference between the left- and right-hand sides of Eq.(\ref{eq:Aij}), squared and summed over all pairs $(i, j)$ in 8 sets of $A_{ij}$ and $\Delta\omega$.. With the real part calculated, the imaginary part is subsequently obtained through further minimizing the cost function. Fig.\ref{fig:rho matrix}(b) shows the matrix elements of the imaginary part, which are close to but oscillating around 0. Accordingly we calculate $Tr(\rho_{TM}^2)=92.5\%$, which is defined as the temporal purity of the single-photon state, which agrees with the results reported before using Hong-Ou-Mandel interference\cite{Qian}.

For these subnatural-linewidth single photons, whose coherence time is far longer than the time resolution of SPCM, it is safe to assume that the photonic temporal state is well described by Eq. (\ref{eq:heralded single state time}). The square of magnitude of the TMF $|\varphi|^2$ is directly from the diagonal elements of $Re[\rho_{TM}]$, as shown in Fig.\ref{fig:rho matrix}(c). The phase function $\theta(\tau)$ is evaluated from $tan\theta_j=Im[\rho_{mj}]/Re[\rho_{mj}]$ where $m=50$ is taken in Fig.\ref{fig:rho matrix}(d). The magnitude of the mode function agrees well with the coincidence measurement reported elsewhere before. More importantly, the phase function over $\tau$ is obtained close to 0 within the coherence window of 200 ns, which is well larger than the coherence time corresponding to the natural-linewidth of $^{85}$Rb atoms. This is predicted by the model of EIT assisted SFWM\cite{DuJOSAB}, but not yet proved from coincidence measurement reported before. In particular, the optical precursor\cite{DuPRL,Shanchao} signifying abrupt dispersion is revealed clearly in the magnitude and the corresponding phase is found to be not changed. On the other hand, the Fock state property of the tested single-photon state is verified through measuring the conditional auto-correlation function $g^{(2)}_c=0.34\pm0.01$ from photon counting approach.


\begin{figure*}
\centering
\includegraphics[width=18cm]{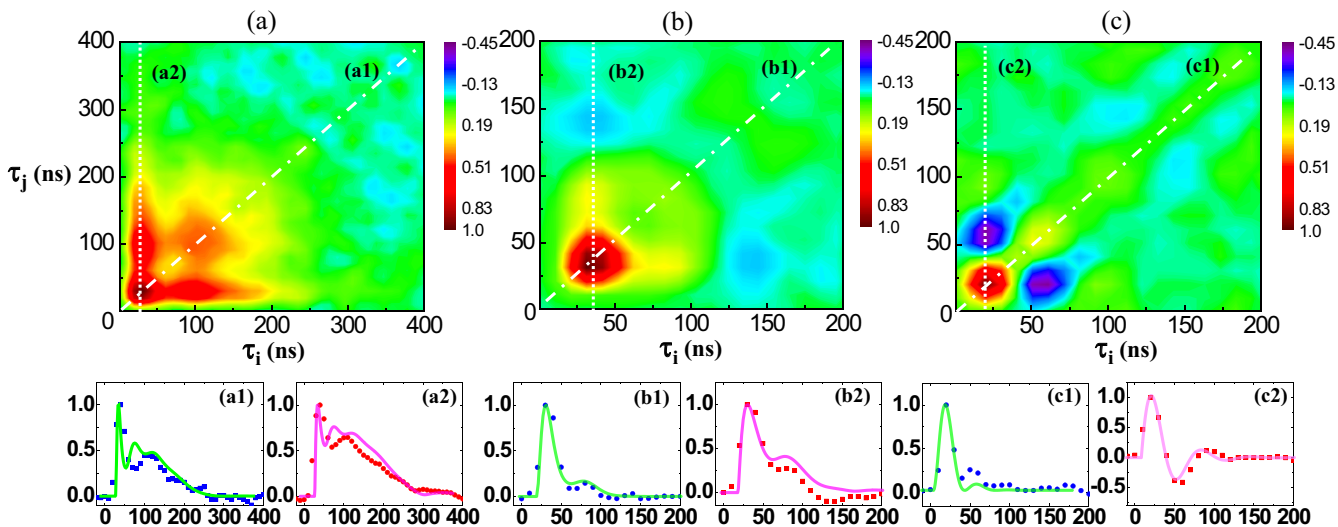}
\caption{Reduced autocorrelation matrix measured from homodyne detection. (a) $OD=53$, $\Omega_c=2\pi\times16.5$ MHz; (b) $OD=25$, $\Omega_c=2\pi\times18$ MHz; (c) $OD=10$, $\Omega_c=2\pi\times31.5$ MHz. The elements along the diagonal axis are plot in (1), while the elements along a vertical axis which cross the maximum value are plot in (2). The solid lines in the plots from (a1) to (c2) are the theoretical curves according to the listed experimental parameters. The maximum element in every matrix is normalized to be 1. The time resolution $\delta\tau$ is 10 ns.}
\label{fig:changing TM mode}
\end{figure*}

Instead of doing the full tomography on the temporal density matrix, we now demonstrate that the homodyne detection is applicable to show the TMF of subnatural-linewidth single photons. As is discussed above, the reduced autocorrelation matrix $A_{ij}$ is equivalent to the real part of the temporal density matrix when $\Delta\omega=0$, and when $\tau_i$ is kept as a constant $m$, from Eq.(\ref{eq:Aij}) the reduced autocorrelation function is,
\begin{eqnarray}
A_{jj}&=&|\varphi(\tau_j)|^2  \\
A_{mj}&=&\varphi^{\ast}(\tau_m)Re[\varphi(\tau_j)]
\label{eq:Aij2}
\end{eqnarray}

\noindent The matrix elements in Eq.~\ref{eq:Aij2} are taken along a diagonal and vertical axis (passing through the maximum) of the reduced autocorrelation matrix. In Fig.\ref{fig:changing TM mode}, we show three typical temporal modes of the heralded single photons generated. When EIT plays a significant role in the SFWM process, the TMF does not change its phase throughout the coherence time, as is shown by Fig.\ref{fig:changing TM mode}(a). By reducing the optical depth of the cloud while increasing $\Omega_c$, the SFWM process transforms from group delay regime to Rabi oscillation regime. Fig.~\ref{fig:changing TM mode}(b) shows a damped Rabi oscillation. Fig.\ref{fig:changing TM mode}(c) shows the TMF in Rabi oscillation regime, where the double SFWM channels due to AC-Stark shift induced by the coupling field become significant. Beating between photon generation channels leads to a $\pi$ phase change for the mode function. The plots (a1)-(c2) agree well with the theoretical calculation of the biphoton wavefunctions\cite{DuJOSAB,FOP}. In particular, for Rabi oscillation regime, plot (c2) perfectly matches the TMF of $\varphi(\tau)\sim e^{-\gamma_e\tau}sin(\frac{\Omega_e}{2}\tau)$, and plot (c1) matches $|\varphi(\tau)|^2$. Here $\gamma_e=(\gamma_{13}+\gamma_{12})/2$ and $\Omega_{e}=\sqrt{\Omega_c^2-(\gamma_{13}-\gamma_{12})^2}$, where $\gamma_{\mu\nu}$ denotes the dephasing rate between states $\mu$ and $\nu$. Compared to homodyne detection approach, coincidence measurement from photon counting merely gives the plots (a1)-(c1).



For the heralded single photons of subnatural-linewidth, its complete temporal mode is revealed in the complex temporal density matrix. In this work, we demonstrate the temporal density matrix of an EIT reshaped single photon through its heterodyne detection with a strong LO field. The amplitude and phase of the mode function are obtained accordingly. While the amplitude of the mode function agrees well with the coincidence measurement through photon counting technique, the phase indicates to be a constant throughout the coherence time of 200 ns. Different from tomography on the temporal-spectral mode of ultrafast single photons generated from SPDC, tomography introduced in our work is in the basis of time bins. The corresponding purity of the temporal density matrix is given to be 92.5\%, which indicates that an almost intact pure state is maintained in this narrowband photon source. Furthermore, we demonstrate that homodyne detection is a direct technique to show the TMF of subnatural-linewidth single photons.

Up to now, the generation, manipulation and characterization of the temporal-spectral mode of subnatural-linewidth single photons are experimentally complete. In contrast to utilizing spectral modes of ultra-fast single photons from SPDC, the temporal modes of subnatural-linewidth single photons are useful in constructing the orthogonal mode functions, e.g., Hermite-Gaussian functions, directly in time domain. Ascribed to the narrow-linewidth, the photonic state is always pure and engineered conveniently even through slow modulators. Secondly, the temporal modes of subnatural-linewidth single photons are applicable to generate time-bin superposition states, e.g., $|1_j,0_k\rangle\pm|0_j,1_k\rangle$ or $|1_j,0_k\rangle\pm i|0_j,1_k\rangle$ ($j$,$k$ denotes different time-bins). The time-bin qubits are easy to be generalized to multi-dimensional qudits without insertion of beam splitters into Franson interferometer. This is an unexploited degree of freedom to encode photonic information in quantum information science.

\begin{acknowledgements}
This work is supported by the National Key Research and Development Program of China under Grant number 2016YFA0302001, and the National Natural Science Foundation of China through Grant No. 11674100, 11654005, 11234003, the Natural Science Foundation of Shanghai No. 16ZR1448200, and Shanghai Rising-Star Program 17QA1401300.
\end{acknowledgements}


\end{document}